\documentclass[modern]{aastex61}
\usepackage{graphicx}
\usepackage{amsmath}
\usepackage{amssymb}
\usepackage{enumitem}

\newcommand\mybar{\kern1pt\rule[-\dp\strutbox]{.8pt}{\baselineskip}\kern1pt}

\setlist[itemize]{noitemsep, topsep=0pt, leftmargin=*}

\shorttitle{PBH Accretion}
\shortauthors{Loeb}

\begin{document}

\title{Quantum-Mechanical Suppression of Accretion by Primordial
  Black Holes}

\author{Abraham Loeb}
\affiliation{Astronomy Department, Harvard University, 60 Garden
  St., Cambridge, MA 02138, USA}

\begin{abstract}
The Schwarzschild radii of primordial black holes (PBHs) in the mass
range of $6\times 10^{14}~{\rm g}$ to $4\times 10^{19}~{\rm g}$ match
the sizes of nuclei to atoms.  I discuss the resulting
quantum-mechanical suppression in the accretion of matter by PBHs in
dense astrophysical environments, such as planets or stars.
\end{abstract}

\bigskip\bigskip
\section{Introduction}

Current cosmological constraints allow for the possibility that dark
matter is made of primordial black holes (PBHs) in the asteroid mass
range of $\sim 10^{18}$-$10^{22}~{\rm
  g}$~\citep{1974MNRAS.168..399C,2021arXiv211002821C,2024NuPhB100316494G,2024arXiv240605736C}.

The accretion of atomic gas or a plasma onto PBHs was described so
far~\citep{2022PhLB..83237265D,2023arXiv231214097D,2023arXiv230309341F,2024arXiv240304227C}
based on the hydrodynamics of a continuous fluid. Here, I point out
that this formalism is invalid because of quantum-mechanical effects
which dominate for asteroid-mass PBHs.

A classical black hole is the ultimate prison.  However, it is
difficult to fit a plump prisoner into a prison that is much smaller
than the prisoner's body size. In the case of a black hole, the prison
walls are represented by the event horizon, which in the non-spinning
case is a sphere with the Schwarzschild radius,
\begin{equation}
  r_{\rm Sch}=\left({2GM\over c^2}\right)= 5.3\times 10^{-9}~{\rm
    cm}\left({M\over 3.6\times 10^{19}~{\rm g}}\right)~,
\label{Sch}
\end{equation}
where $G$ is Newton’s constant, $c$ is the speed of light and $M$ is
the black hole mass~\citep{1916AbhKP1916..189S}. The Schwarzschild
radius equals the Bohr radius of the hydrogen
atom~\citep{1913Natur..92..231B},
\begin{equation}
r_{\rm B}={\hbar^2 \over m_e e^2}= 5.3\times 10^{-9}~{\rm cm}~,
\end{equation}
for a black hole mass of $M=3.6\times 10^{19}~{\rm g}$.  PBHs of this
mass could have been produced in the early Universe. The Schwarzschild
radius equals the radius of the proton, $r_p=8.4\times 10^{-14}~{\rm
  cm}$~\citep{2013Sci...339..417A} for a black hole mass of
$M=5.7\times 10^{14}~{\rm g}$. This happens to be a few times the mass
of a PBH that evaporates by Hawking
radiation~\citep{1974Natur.248...30H} on a timescale comparable to the
age of the Universe. Much smaller PBHs would have disappeared by now.

\section{Quantum Suppression of Astrophysical Accretion}

Black holes grow in mass by accreting matter from their astrophysical
environment~\citep{2014ARA&A..52..529Y,2014SSRv..183...21B}. For black
holes more massive than the Sun, the event horizon is large enough to
be considered as a mouth that absorbs many atoms at once and so the
accreted matter can be approximated as a continuous fluid. But in the
regime of asteroid-mass PBHs, this assumption is no longer
valid. These black holes feast on a single atom, a single proton or a
single electron at a time, because their event horizon is smaller than
the spatial extent of the quantum-mechanical wave function of these
particles.

Consider a situation where a hydrogen atom is attracted
gravitationally towards a PBH. In the naive perception of the atom as
a point particle, it could reach the black hole center from a distance
$r$ over a free-fall time, $\sim (r^3/GM)^{1/2}$. However, if the
horizon is smaller than the size of the atom, then this `plump
prisoner' will not be captured and most of the atom will stay
predominantly outside the event horizon. Even if the PBH is massive
enough to absorb the more compact proton, the extended wave function
of the electron could remain outside the PBH. In that case, the PBH
will acquire a positive electric charge. The electric force that binds
the electron to the proton will be enhanced by the PBH gravity.

At distances much larger than the horizon, both electric and
gravitational forces decline inversely with distance squared. Hence,
the effective Bohr radius (marked hereafter by a tilde) of the {\it
  hydrogen$+$PBH} atom shrinks to a value,
\begin{equation}
{\tilde r}_{\rm B}={\hbar^2 \over m_e (e^2+GMm_e)}= 5.3\times
10^{-9}~{\rm cm}\times \left[1+ \left({M\over 3.8\times 10^{15}~{\rm
      g}}\right)\right]^{-1}~.
\label{Bohr}
\end{equation}
Correspondingly, the effective Rydberg energy levels will be modified
to $E_n=-1{\rm {\tilde Ry}}/n^2$, with $n$ an integer. The
modified binding energy of the ground state would be,
\begin{equation}
-E_1= 1{\rm {\tilde Ry}}={\hbar^2 \over 2 m_e {\tilde r}_{\rm B}^2}=
13.6~{\rm eV}\times \left[1+ \left({M\over 3.8\times 10^{15}~{\rm
      g}}\right)\right]^{2}~.
\label{BohrE}
\end{equation}

The gravitational and electric forces are of equal magnitude for a PBH
mass of $M=3.8\times 10^{15}~{\rm g}$, and gravity wins at larger
masses.  The atomic radius in equation (\ref{Bohr}) shrinks below the
Schwarzschild radius in equation (\ref{Sch}) for a PBH mass
$M=3.7\times 10^{17}~{\rm g}$. At this mass, the gravitational binding
energy in equation (\ref{BohrE}) approaches a quarter of the electron
rest-mass energy. For higher PBH masses, the Compton wavelength of the
electron, $(\hbar/m_e c)=4\times 10^{-11}~{\rm cm}$, is smaller than
the Schwarzschild radius, and the ground state of the bound electron
can be inside the horizon. The criterion that accretion of an electron
is suppressed is identical, up to a numerical factor of order unity,
to the condition that electron-positron pairs are emitted through
Hawking evaporation of the black hole.

For lower mass PBHs, accretion could be significantly suppressed in
common astrophysical environments, where gas particles have
non-relativistic thermal speeds and the mass density is much lower
than the nuclear density of neutron stars~\citep{Giffin}. In the
rarefied environments of the interstellar or intergalactic media, a
low-mass PBH would accrete one particle at a time with a significant
quantum-mechanical suppression in accretion rate relative to
hydrodynamic expectations.

\section{Accretion Rate}

To obtain reliable assessments of the accretion rate by PBHs with
masses $M\lesssim 4\times 10^{19}~{\rm g}$, one must calculate the
wave function of electrons and protons using the Dirac equation in the
background PBH metric. For each quantum state and energy level, the
overlap of the electron wave function with the volume interior to the
event horizon sets a finite half-life for the electron to stay in a
bound state outside the horizon. Afterwards, the electron will join
the proton inside the horizon and neutralize the PBH charge.

In the quantum world, there is a finite probability per unit time for
a plump atom to be captured by a small event horizon. The quantum
transition to the final state of capture resembles tunneling through a
barrier.

\subsection{Rarefied Cold Gas with Bound Electrons}

The accretion of particles in bound states around a PBH or in tightly
packed systems is different from absorption of free particles by a
PBH~\citep{Unruh}.  A bound electron can be captured by an atomic
nucleus even without a central black hole. This well-known process,
called electron capture, involves the absorption an electron from an
inner atomic shell by a proton-rich nucleus of a neutral
atom~\citep{1937PhRv...52..134A}. The rate for this capture is related
to the overlap of the electron wave function with the volume of the
atomic nucleus.

Similarly, the accretion of protons and electrons by a PBH in the
asteroid mass range, is dictated by the overlap of their
quantum-mechanical wave function with the volume interior to the PBH
event horizon.

Ignoring Hawking radiation, an estimate for the half-life of a bound
electron around a black hole with $r_{\rm Sch}<{\tilde r}_{\rm B}$ (or
equivalently $M< 3.7\times 10^{17}~{\rm g}$) which already captured a
proton, is
\begin{equation}
\tau_{1/2}\sim {({\tilde r}_{\rm B}/{\tilde v}_{\rm B})\over (r_{\rm
    Sch}/{\tilde r}_{\rm B})^3}= 2.2\times 10^{-11}~{\rm s}
\left({M\over 3.8\times 10^{15}~{\rm g}}\right)^{-3} \left[1+
  \left({M\over 3.8\times 10^{15}~{\rm g}}\right)\right]^{-5}~,
\label{tau}
\end{equation}
where the effective electron speed is,
\begin{equation}
  {\tilde v}_{\rm B}\sim \left[{(e^2+GMm_e)\over \hbar c}\right] c =
  2.2\times 10^{8}~{\rm cm~s^{-1}} \left[1+ \left({M\over 3.8\times
      10^{15}~{\rm g}}\right)\right]~.
\label{velB}
\end{equation}
The accretion rate corresponding to the absorption of an electron-proton pair
per $\tau_{1/2}$ is,
\begin{equation}
  {\dot M}\equiv {m_p\over \tau_{1/2}}=7.5\times 10^{-14}~{\rm g~s^{-1}}~
  \left({M\over 3.8\times
    10^{15}~{\rm g}}\right)^{3} \left[1+ \left({M\over 3.8\times
      10^{15}~{\rm g}}\right)\right]^{5}~.
  \label{mdot}
\end{equation}
The accretion rate is too low to add significant mass to PBHs with
$M\lesssim 1.4\times 10^{17}~{\rm g}$ over the entire age of the
Universe.

At high enough infall rate, a bound state could involve multiple
electrons and protons simultaneously. When the inflow of fresh protons
into the PBH exceeds $\tau_{1/2}^{-1}$, the PBH could be charged
positively up to a maximum charge, $Q_{\rm max}\sim (k_{\rm
  B}T/m_p)(GMm_p/e^2)e=(T/1.1\times 10^{13}~{\rm K})(M/2.1\times
10^{12}~{\rm g})e$, at which the electric potential near the horizon
equals the thermal energy for external protons at the background
temperature~$T$~\citep{Med}. At this maximum charge (which will be
screened over the Debye length in the surrounding plasma), the
accretion rate can be enhanced relative to equation~(\ref{mdot}) since
the effective Bohr radius for the ground state of the innermost
electron at $Q_{\rm max}$ shrinks below the Schwarzschild radius for
PBH masses $M\gtrsim 8.6\times 10^{15}~{\rm g}$. However, at the high
plasma temperatures expected near the PBH horizon, the electrons will
not stay in bound states and the accretion rate will be dictated by
the dynamics of free particles.

\subsection{Dense Hot Plasma with Free Electrons}

For a thermal plasma of free (unbound) particles, the accretion rate
can exceed the value expected from equation~(\ref{mdot}) if the
interparticle distance $d=(4\pi n_e/3)^{-1/3}$ is shorter than
${\tilde r}_{\rm B}$, where $n_e$ is the electron number density in
the plasma. In this regime, the fastest accretion will involve one
electron-proton pair per $\tau_{1/2}$, where the length scale ${\tilde
  r}_{\rm B}$ in equation~(\ref{tau}) is replaced by the interparticle
distance $d$ and the velocity scale ${\tilde v}_{\rm B}$ is replaced
by the thermal speed of the particles $v_\infty$ far from the PBH
(assuming a stationary PBH relative to the plasma rest frame). This
substitution gives a net accretion rate of,
\begin{equation}
{\dot M} = {4\over 3}\left({r_{\rm Sch}\over d}\right)
\rho_\infty v_\infty\left(\pi r_{\rm Sch}^2\right)~,
\label{dense}
\end{equation}
where $\rho_\infty=m_p n_e$ is the plasma mass density far from the
PBH.

It is instructive to compare this suppressed accretion rate to the
Bondi accretion rate~\citep{Bondi}, derived in the fluid
approximation,
\begin{equation}
{\dot M}_{\rm Bondi} \approx \left({c \over v_\infty}\right)^2 \rho_\infty
  v_\infty\left(\pi r_{\rm Sch}^2\right) ~.
\label{dense}
\end{equation}
The resulting suppression factor for $d>r_{\rm Sch}$ is,
\begin{equation}
F_{\rm supp}=\left({{\dot M} \over {\dot M}_{\rm Bondi}}\right)
\approx \left({v_\infty\over c}\right)^2\left({4\over 3}{r_{\rm
    Sch}\over d}\right) ~.
\label{supp}
\end{equation}
For $r_{\rm Sch}\ll d$ and $v_\infty\ll c$, we find that $F_{\rm
  supp}\ll 1$.

This result applies to high matter densities at which atoms are
densely packed and $d< \tilde{r}_{\rm B}$. Such densities are relevant
when PBHs pass through the interiors of planets or
stars~\citep{Hawking,Pretorius,Caplan}, with the most dense medium realized
for PBHs trapped in the interior of a neutron
star~\citep{Pani,Giffin}. For context, the Bondi accretion rate of a
trapped PBH at the mean density of a white dwarf is substantial for
$v_\infty<5\times 10^3~{\rm km~s^{-1}}$,
\begin{equation}
{\dot M}_{\rm Bondi}\approx 1.5\times 10^5~{\rm g~s^{-1}}\left({\rho_\infty\over 10^6~{\rm g~cm^{-3}}}\right)\left({M\over
  3.6\times 10^{19}~{\rm g}}\right)^2\left({v_\infty\over 5\times 10^3~{\rm
      km~s^{-1}}}\right)^{-3}.
\end{equation}
Expressing this equation as ${\dot M}=A M^2$ with $A$ being a
constant, yields a solution for the final PBH mass $M_f$ in terms of
its initial mass $M_i$ as, $M_f=1/(1/M_i-At)$. This solution implies
that over a timescale shorter than 10~Myr, the mass of a trapped PBH
with $M_i= 3.6\times 10^{19}~{\rm g}$, would diverge through Bondi
accretion well above its initial value. Consequently, a trapped PBH
could have had dramatic observable consequences on the energy budget
or the ultimate fate of a white dwarf. However, in reality the PBH
growth would be suppressed by the quantum-mechnical effects discussed
above as well as by the Eddington limit on the accretion rate, which
is $\sim 3\times 10^5~{\rm g~s^{-1}}(\epsilon/0.01)^{-1}(M/3.6\times
10^{19}~{\rm g})$, where $\epsilon$ is the radiative efficiency.

Other dense environments involve the thermal electron-positron plasma
of the early universe~\citep{2024arXiv240304227C}, where the thermal
speed of the particles is $\sim c$. The de Broglie wavelength of
relativistic electrons and positrons in this thermal plasma of free
particles is of order $d$, smaller than their Compton wavelength. This
is properly incorporated in the suppression factor of
equation~(\ref{supp}) which is effective for $d>r_{\rm Sch}$.  For a
wavefunction with a de Broglie wavelength larger than the
Schwarzschild radius, the relevant momentum is calculated far from the
horizon, where most of the probability distribution of finding the
particle resides.

\section{Concluding Remarks}

Including Hawking radiation~\citep{1974Natur.248...30H} would further
lower the accretion rate since the Hawking temperature, $T_{\rm
  H}=(\hbar c^3/8\pi k_BGM)=2.7\times 10^6~{\rm eV}~(M/3.8\times
10^{15}~{\rm g})^{-1}$, exceeds the binding energy $1{\rm {\tilde
    Ry}}$ in equation~(\ref{BohrE}), and the Hawking luminosity,
$L_{\rm H}=(\hbar c^6/15360\pi G^2M^2)=1.5\times 10^{26}~{\rm
  eV~s^{-1}}~(M/3.8\times 10^{15}~{\rm g})^{-2}$, exceeds $1{\rm
  {\tilde Ry}}/\tau_{1/2}$. The outward flux of high energy photons
and electron-positron pairs could suppress accretion altogether.

The accretion of matter into asteroid-mass PBHs is only significant in
the dense interiors of planets or stars, but can be ignored at the
much lower cosmic-mean-density of matter and radiation in the
Universe~\citep{Zhang}.

If we ever witness an asteroid-mass black hole in the solar system, it
could serve as a testbed for quantum-gravitational physics on a
subatomic scale.

\bigskip
\bigskip
\bigskip
\bigskip
\section*{Acknowledgements}

I am grateful for insightful comments from an anonymous referee. This
work was supported in part by Harvard's {\it Black Hole Initiative},
which is funded by grants from JFT and GBMF.

\bigskip
\bigskip
\bigskip

\bibliographystyle{aasjournal}
\bibliography{t}
\label{lastpage}
\end{document}